\documentclass[preprint,showpacs,preprintnumbers,amsmath,amssymb]{revtex4}

\usepackage{graphicx}
\usepackage{epsfig}		
\usepackage{dcolumn}
\usepackage{bm}

\def\0{\mbox{\tiny $0$}}
\def\1{\mbox{\tiny $1$}}
\def\2{\mbox{\tiny $2$}}
\def\3{\mbox{\tiny $3$}}
\def\4{\mbox{\tiny $4$}}
\def\5{\mbox{\tiny $5$}}
\def\6{\mbox{\tiny $6$}}
\def\7{\mbox{\tiny $7$}}
\def\8{\mbox{\tiny $8$}}
\def\9{\mbox{\tiny $9$}}
\def\R{\mbox{\tiny $\mathit{R}$}}
\def\L{\mbox{\tiny $\mathit{L}$}}
\def\T{\mbox{\tiny $T$}}
\def\A{\mbox{\tiny $\alpha$}}
\def\B{\mbox{\tiny $\beta$}}
\def\bbb{\mbox{\tiny $B$}}
\def\bb#1{\mbox{\footnotesize $(#1)$}}

\def\I{\mbox{\tiny $I$}}
\def\II{\mbox{\tiny $II$}}
\def\III{\mbox{\tiny $III$}}

\def\n{\mbox{\tiny $n$}}

\def\f14{\mbox{\tiny $\frac{1}{4}$}}
\def\mi{\mbox{\tiny $-$}}
\def\pl{\mbox{\tiny $+$}}

\begin{document}

\title{Recovering the stationary phase condition for accurately obtaining scattering and tunneling times}

\author{A. E. Bernardini}
\email{alexeb@ifi.unicamp.br}
\affiliation{Instituto de F\'{\i}sica Gleb Wataghin, UNICAMP,
PO Box 6165, 13083-970, Campinas, SP, Brasil.}

\date{\today}







\begin{abstract}
The stationary phase method is often employed for computing tunneling {\em phase} times of analytically-continuous {\em gaussian} or infinite-bandwidth step pulses which collide with a potential barrier.
The indiscriminate utilization of this method without considering the barrier boundary effects leads to some misconceptions in the interpretation of the phase times.
After reexamining the above barrier diffusion problem where we notice the wave packet collision necessarily leads to the possibility of multiple reflected and transmitted wave packets, we study the phase times for tunneling/reflecting particles in a framework where an idea of multiple wave packet decomposition is recovered.
To partially overcome the analytical incongruities which rise up when tunneling phase time expressions are obtained, we present a theoretical exercise involving a symmetrical collision between two identical wave packets and a one dimensional squared potential barrier where the scattered wave packets can be recomposed by summing the amplitudes of simultaneously reflected and transmitted waves.
\end{abstract}


\maketitle

\section{INTRODUCTION}

\hspace{1 em} The analytical methods utilized for reproducing the wave packet collision with a potential barrier have been widely discussed in the context of (one dimensional) scattering and tunneling phenomena \cite{Hau89,Lan94,Per01}.
In this extensively explored scenario, the stationary phase method (SPM) is the simplest and the most common approximation employed for obtaining the group velocity of a wave packet in a quantum scattering process which represents the collision of a particle with a square (rectangular) barrier of potential.
First introduced to physics by Stokes and Kelvin \cite{K887}, the SPM essentially provides an approximate value for the maximum of an integral in an analytical procedure which enables us
to understand several subtleties of interesting quantum phenomena, such as tunneling \cite{Hau89,Ste93,Bro94}, resonances \cite{Bra70}, incidence-reflection and incidence-transmission interferences \cite{Per01} as well as the Hartman effect \cite{Har62} and the correspondent {\em superluminal} transition time interpretation \cite{Olk92,Hay94,Jak98}.
It is also often quoted in the context of testing different theories for temporal quantities such as arrival, dwell and delay times \cite{Hau89,Lan94,Win03} and the asymptotic behavior at long times \cite{Jak98,Bau01}.

The study presented in this manuscript concerns with some limitations on the use of the SPM for obtaining the right position of a propagating wave packet subject to a very simple and common potential configuration.
The limitations are not concerned with the domain of the method, i. e. with the mathematical basis and the analytical applicability of the method, they focus on the indiscriminate application of the method to time-dependent scattering problems and, in particular, for deriving phase times.
The time-dependent scattering of a wave-packet by a potential barrier in the linear Schroedinger equation is interesting because it reconciles two types of wave theories.
Those interested in linear theory often look at stationary (frequency-domain) solutions of a single frequency, whereas those interested in nonlinear theory usually look at time-dependent pulses.
To implement the accurate analysis with the SPM, we work with the hypothesis of adding up the Fourier modes from the stationary solution to observe the evolution of a linear wave packet on collision with an obstacle.
In the first part of the manuscript, the analysis of the applicability of the SPM is concentrated on the investigation of the non-relativistic diffusion of an incoming single wave packet with energy spectrum totally localized above the potential barrier.
We are particularly interested in obtaining an analytical description of the above barrier diffusion problem by pointing out some interpretive problems which concern with non-conservation of probability during the collision process.
In the second part of our analysis, the correct quantification of the analytical incongruities which restrict the applicability of the principle of stationary phase is extended to the study a tunneling particle where, eventually, the {\em superluminal} interpretation of transition times can be ruined.
In order to organize the ideas which are discussed in the manuscript, we have stated the following division.
In section II we discuss the general applicability of the SPM in a wave packet propagation problem by introducing an analytical description of the above barrier diffusion problem where we reobtain the solution for the apparent physical inconsistency which concerns with the non-conservation of probability during the collision process.
By recurring to a wave packet multiple peak decomposition \cite{Ber04}, we can decompose the original colliding wave packet into {\em multiple} reflected and transmitted components by means of which the conservation of probabilities is recovered during the scattering process.
The analytic expressions for all the reflected, transmitted and intermediary components are obtained and the validity of them is discussed in terms of an illustrative example where the analytic approximation and the reflects exactly the numerical results.
The physical framework where the study of a tunneling process is embedded is permeated by theoretical constructions presented in section III.
It involves analytically-continuous {\em gaussian} pulses which due to its not well-defined front lead to ambiguities in the interpretation of the wave packet speed of propagation.
Beside it, infinite bandwidth signals cannot propagate through any real physical medium (whose transfer function is therefore finite) without pulse distortion, which also leads to ambiguities in determining propagation velocity during the tunneling process.
Moreover, some of the barrier transposing time definitions lead, in tunneling time conditions, to very short times which can even become negative where they may seem to contradict simple concepts of causality.
To partially overcome these incompatibilities, in section IV, we discuss a theoretical exercise involving a symmetrical collision with a one dimensional square potential where we recompose the scattered wave packet by summing the amplitudes of the reflected and transmitted waves in the scope of what we classify as a multiple peak decomposition analysis \cite{Ber04} where the conditions for the SPM applicability are totally recovered.
Finally, we draw our conclusions in section V.

\section{THE ABOVE BARRIER DIFFUSION PROBLEM}

The principle of stationary phase to describe the group velocity is well-established, and was employed by Sommerfeld and Brillouin in their early studies of wave propagation and group velocity \cite{Bri60}.
The method has in time become a standard tool for several theoretical applications \cite{PBE} adopted not only by physicists but also by biologists, economists etc.
It can also be used in a statistical sense, such that the most likely events tend to be associated with slowly-varying phase variation in the frequency domain, and unlikely  events tend to be associated with rapidly varying phase with frequency.

To illustrate the applicability of the SPM into a physical problem it should be instructive to assume an equivalence between the complex variable $z$ and a real one $k$ so that an integral which represents a free quantum particle propagation can be identified by the wave packet solution of the unidimensional Schroedinger equation,
\small\begin{eqnarray}
i\hbar\frac{\partial}{\partial t} \psi(x, t)  &=& \left[-\frac{\hbar^{\2}}{2 m} \frac{d^{\2}}{dx^{\2}} + V\bb{x}\right] \psi(x, t)
\label{201b}
\end{eqnarray}\normalsize
with $V\bb{x} = 0$, described in terms of the integral,
\small\begin{eqnarray}
\psi(x, t)  &=& \int_{_{-\infty}}^{^{+\infty}}\frac{d k}{\sqrt{2\pi}}\,
G(k , k_{\0}) \, \exp{\left[- i\, (E\bb{k} \, t +  k \, x - k \, x_{\0})\right]},
\label{201}
\end{eqnarray}\normalsize
where we have set $\hbar = 1$ and the dispersion relation as $E\bb{k} = k^{\2}/(2m)$.
The integral (\ref{201}) can therefore be estimated by finding the place where the phase has a vanishing derivative, evaluating (approximately) the integral in the neighborhood of this point.
The movement of the peak coordinate of the wave packet $\psi(x,t)$ can be obtained by imposing the stationary phase condition
\small\begin{equation}
\left.\frac{d}{dk}[E \, t - \, k (x - x_{\0})]\right|_{k = k_{\0}} = 0
~~~~\Rightarrow~~~~x_{\mbox{\tiny max}} = x_{\0} + \frac{k_{\0}}{m} \, t,
\label{202}
\end{equation}\normalsize
which means that the maximum of the wave packet propagates with a velocity
$v = \frac{k_{\0}}{m}$.
In fact, we ratify this result when we explicitly calculate the integral of
Eq.~(\ref{201}) by introducing a {\em gaussian} momentum distribution given by
\small\begin{equation}
G(k , k_{\0}) = g(k - k_{\0}) = \left(\frac{a^{\2}}{2 \, \pi}\right)^{\frac{1}{4}}
\exp{\left[-\frac{a^{\2} (k -k_{\0})^{\2}}{4}\right]}.
\label{203}
\end{equation}\normalsize
In this case, the result of the integration (\ref{201}) give us
\small\begin{equation}
\psi(x,t) = \varphi[x - x_{\0}, t],
\label{204A}
\end{equation}\normalsize
where
\small\begin{equation}
\varphi[x, t] = \left[ \mbox{$\frac{\,\,\,\pi a^{\2}}{2}$} \,
\left( 1 + \mbox{$\frac{4\,t^{\2}}{m^{\2}a^{\4}}$} \right)
\right]^{-\frac{1}{4}} \exp \left[ -
\frac{\left(x-\frac{k_{\0}}{m}\,t\right)^{\2}}{a^{\2}\left(
1+\frac{2\,i\,t}{ma^{\2}}\right)}-\mbox{$\frac{i}{2}$} \arctan
\mbox{$\left(\frac{2t}{ma^{\2}}\right)$}+i (k_{\0}x - E_{\0}t) \right]
\label{204}
\end{equation}\normalsize
which evidently ratifies the result of Eq.~(\ref{202}).

Meanwhile, the method leads to some new interpretive discussion when the momentum distribution becomes a complex function, i. e. when $G(k , k_{\0})$ has to be written as $|G(k, k_{\0})|\, \exp{[i\,\lambda(k)]}$.
In this case, the stationary phase condition for a free wave packet represented by Eq.~(\ref{202}) will be modified by the presence of an additional phase $\lambda(k)$.
By assuming that $\lambda(k)$ can be expanded in the neighborhood of $k_{\0}$, i. e.
\small\begin{equation}
\lambda(k) \approx \lambda(k_{\0}) + (k - k_{\0})
\left.\frac{d \lambda(k)}{dk}\right|_{k = k_{\0}}
\label{205}
\end{equation}\normalsize
the new stationary phase condition should give
\small\begin{equation}
x_{\mbox{\tiny max}} = x_{\0} - \left.\frac{d \lambda(k)}{dk}\right|_{k = k_{\0}}
+ \frac{k_{\0}}{m} \, t.
\label{206}
\end{equation}\normalsize
As it is commonly presented in several textbooks of quantum mechanics \cite{Coh77}, one can be persuaded to take the stationary phase condition as a necessary and {\em sufficient} statement for applying the SPM.
Nevertheless, although the principle of stationary phase can be used to predict the most likely outcome of an experiment, it does not exclude the possibility of alternative outcomes.
In fact, the results of the SPM depend critically upon the manipulation of the generic amplitude $G(k,k_{\0})$ prior to the application of the method.
To clear up this point, let us firstly analyze the diffusion of a wave packet by a squared potential $V_{\0}$ when the incoming energies are higher than $V_{\0}$.
The stationary wave solution of the Schroedinger equation (\ref{201b}) computed by considering the potential barrier described by
\small\begin{equation}
\begin{array}{clll}  0&&~~~\mbox{if}~~~x < 0&~~~~\mbox{region I}, \\
 V_o &&~~~\mbox{if}~~~0 < x < L&~~~~\mbox{region II}, \\
  0 &&~~~\mbox{if}~~~x > L&~~~~\mbox{region III},
\end{array}
\label{p2}
\end{equation}\normalsize
can be decomposed into different wave functions for each interval of $x$, i. e.
\small\begin{equation}
\begin{array}{ccccccc}
\Phi(k,x) &=& \phi_{\I}(k, x) &  + &
                \phi_{\II}(k, x) & +  &\hspace*{.8cm}
                \phi_{\III}(k, x), \\
& & \overbrace{\phi_{Inc}(k, x) + \phi_{\R}(k, x)} &  &
                \overbrace{\phi_{\A}(k,x) +\phi_{\B}(k, x)} &  &
                 \hspace*{.8cm} \overbrace{\phi_{\T}(k, x)}
\end{array}
\label{p3}
\end{equation}\normalsize
where
\small\begin{eqnarray}
\phi_{Inc}(k, x) &=& \exp{\left[ i \,k \,x\right]},\nonumber\\
\phi_{\R}(k, x)  &=& R(k)\exp{\left[ - i \,k \,x\right]},\nonumber\\
\phi_{\A}(k, x)  &=& \alpha(k)\exp{\left[ i \,q  \,x\right]},\nonumber\\
\phi_{\B}(k, x)  &=& \beta(k)\exp{\left[ - i \,q  \,x\right]},\nonumber\\
\phi_{\T}(k, x)  &=& T(k)\exp{ \left[ i \,k \,x\right]},
\label{p4}
\end{eqnarray}\normalsize
with $q = (k^{\2} - w^{\2})^{\frac{1}{2}}$ and $w = \sqrt{2\, m \, V_{\0}}$.
By solving the constraint equations at $x = 0$ and $x = L$
we obtain
\small\begin{eqnarray}
\alpha(k) &=&     \left[\frac{k(k + q)}{\mathcal{F}(k)}\right]
\exp{\left[i\,\Theta(k) - i\,q \, L\right]},\nonumber\\
\beta(k)  &=& -   \left[\frac{k(k - q)}{\mathcal{F}(k)}\right]
\exp{\left[i\,\Theta(k) + i\,q \, L\right]},\nonumber\\
R(k)      &=&   -i \,  \left[\frac{k^{\2} - q^{\2}}{\mathcal{F}(k)}\right]\sin{[q \,L]}
\exp{\left[i\,\Theta(k)\right]},\nonumber\\
T(k)      &=&     \left[\frac{2 \, k \, q }{\mathcal{F}(k)}\right]
\exp{\left[i\,\Theta(k) - i\,k\, L\right]},
\label{p5}
\end{eqnarray}\normalsize
where
\small\begin{eqnarray}
\mathcal{F}(k) &=& \left\{4 \, k^{\2} \, q^{\2} \cos^{\2}{[q \,L]} +
\left(k^{\2} + q^{\2} \right)^{\2}\sin^{\2}{[q \,L]}\right\}^{\frac{1}{2}}
\end{eqnarray}\normalsize
and
\small\begin{eqnarray}
\Theta(k) &=& \arctan{\left\{\frac{k^{\2} + q^{\2} }{2 \, k \,q }\tan{[q \,L]}\right\}}.
\label{p6}
\end{eqnarray}\normalsize
The explicit expression for the correspondent propagating wave packets can be obtained by solving the integrals like
\small\begin{eqnarray}
\psi_f(x, t)  &=& \int_{_{w}}^{^{+\infty}}\frac{d k}{\sqrt{2\pi}}\,
g(k - k_{\0}) \, \phi_f(k,x)\,\exp{[- i\,E \, t]},
\label{p6A}
\end{eqnarray}\normalsize
with $f \equiv \alpha,\,\beta,\, R,\,T$.
As first approximation which is commonly used in quantum mechanics textbooks \cite{Coh77}, we obtain the analytical formulas to these integrals by assuming the momentum distribution $g(k - k_{\0})$ is sufficiently sharply peaked around a maximum point $k_{\0} > w$.
In this case, the integration can be extended from $[w, \infty]$ to $[-\infty,\infty]$ without modifying the final result.
However, in the sense we are investigating, as we shall demonstrate in the following, it is {\em erroneously} assumed that the $k$-dependent phase terms $\Theta(k)$ and $q$ can be approximately described by a series expansion around $k = k_{\0}$ up to the first order term, i. e.
\small\begin{equation}
\Theta(k) \approx \Theta(k_{\0}) + (k - k_{\0})\Theta^{\prime}(k_{\0})
~~~~\mbox{and}~~~~
q  \approx q_{\0}  + (k - k_{\0})\,
\left.\frac{d q }{d k}\right|_{k = k_{\0}}
\label{p6B}
\end{equation}\normalsize
where
\small\begin{equation}
\Theta^{\prime}(k)
 = \frac{2\,m}{q}
\left[\frac{\left(k^{\2} + q^{\2} \right)k^{\2} \,q \,L - \left(k^{\2} - q^{\2} \right)^{\2}
\sin{[q \,L]}\cos{[q \,L]}}{4 \, k^{\2} \, q^{\2}
+ \left(k^{\2} - q^{\2} \right)^{\2}\sin^{\2}{[q \,L]}}\right]
~~~~\mbox{and}~~~~
\left.\frac{d q }{d k}\right|_{k = k_{\0}} =
 \frac{k_{\0}}{q_{\0}}.
\label{p10}
\nonumber\\
\end{equation}\normalsize
At the same time, when $k$ is approximated by $k_{\0}$, it is assumed that the modulating amplitude $|\phi_f(k,x)|$ can be put out of the integral giving the following results,
\small\begin{eqnarray}
\psi_{Inc}(x, t) &\approx& \varphi[x - x_{\0}, t],\nonumber\\
\psi_{\R}(x, t)  &\approx& R(k_{\0})\,
\int_{_{-\infty}}^{^{+\infty}}\frac{d k}{\sqrt{2\pi}}\,
\mbox{$g(k - k_{\0}) \, \exp{\left[- i\,E \, t - i \,k (x + x_{\0}) +
+ i\, (k - k_{\0})\Theta^{\prime}(k_{\0})
\right]}$}\nonumber\\
& = & \mbox{$R(k_{\0})
\exp{\left[- i \, k_{\0}\,\Theta^{\prime}(k_{\0})\right]}\,
\varphi\left[- x - x_{\0} + \Theta^{\prime}(k_{\0}), t\right]$},\nonumber\\
\psi_{\A}(x, t)  &\approx& \alpha(k_{\0})\,
\int_{_{-\infty}}^{^{+\infty}}\frac{d k}{\sqrt{2\pi}}\,
\mbox{$g(k - k_{\0}) \, \exp{\left[- i \left(E \, t + k \, x_{\0}- q_{\0}\,x\right)
 + i \,(k - k_{\0})\left(\frac{k_{\0}}{q_{\0}}(x- L)
 + \Theta^{\prime}(k_{\0})\right)
\right]}$}\nonumber\\
& = & \mbox{$\alpha(k_{\0})
\exp{\left[i \, q_{\0} \, x - i \, \frac{k^2_{\0}}{q_{\0}}\,(x - L)
 - i \, k_{\0}\Theta^{\prime}(k_{\0})\right]}\,
\varphi\left[\frac{k_{\0}}{q_{\0}}(x - L)
  - x_{\0} + \Theta^{\prime}(k_{\0}), t\right]$},\nonumber\\
\psi_{\B}(x, t)  &\approx& \beta(k_{\0})
\int_{_{-\infty}}^{^{+\infty}}\frac{d k}{\sqrt{2\pi}}\,
\mbox{$g(k - k_{\0}) \, \exp{\left[- i \left(E \, t + k \, x_{\0}+ q_{\0}\,x\right)
- i \, (k - k_{\0})\left(\frac{k_{\0}}{q_{\0}}(x - L)
- \Theta^{\prime}(k_{\0})\right)
\right]}$}\nonumber\\
& = & \mbox{$\beta(k_{\0})
\exp{\left[- i \, q_{\0} \, x + i \, \frac{k^2_{\0}}{q_{\0}}\,(x - L)
 - i \, k_{\0}\Theta^{\prime}(k_{\0})\right]}\,
\varphi\left[-\frac{k_{\0}}{q_{\0}}(x - L)- x_{\0}
 + \Theta^{\prime}(k_{\0}), t\right]$},\nonumber\\
\psi_{\T}(x, t)  &\approx& T(k_{\0})\,
\int_{_{-\infty}}^{^{+\infty}}\frac{d k}{\sqrt{2\pi}}\,
\mbox{$g(k - k_{\0}) \, \exp{\left[- i\,E \, t + i \,k (x - x_{\0}) +
 i\, (k - k_{\0})\left(\Theta^{\prime}(k_{\0})- L\right)
\right]}$}\nonumber\\
& = & \mbox{$ T(k_{\0})
\exp{\left[- i \, k_{\0}\Theta^{\prime}(k_{\0})\right]}\,
\varphi\left[x - x_{\0} - L + \Theta^{\prime}(k_{\0}), t\right]$}.\nonumber\\
\label{p6C}
\end{eqnarray}\normalsize
By observing the {\em gaussian} shape of $\varphi[x, t]$ given by Eq.~(\ref{204}), one can easily (but wrongly!) identify the position of the peak of the wave packets.
The times corresponding to the position $x$ of the incident and reflected wave packet peaks would be respectively given by
\small\begin{equation}
t_{Inc}(x) = \left.\left[\frac{x - x_{\0}}{v_k}\right]\right|_{k = k_{\0}} ~~~~\mbox{and}~~~~
t_{\R}(x) = \left.\left[-\frac{x + x_{\0} - \Theta^{\prime}(k)}{v_k}\right]\right|_{k = k_{\0}}
\label{p9}
\end{equation}\normalsize
where $v_k = \frac{d E}{d k} = \frac{k}{m}$.
Only $x < 0$ is physical in this result since these waves, by definition, lie in region I.
Since the phase of the incoming wave contains only the plane wave factors, i.e. it is devoid of $\Theta(k)$, the incoming peak reaches the barrier at $x=0$ at time $t_{Inc}(0) = -(x_{\0}/v_{k_{\0}})$ (neglecting interference effects).
The presence of the phase term $\Theta^{\prime}(k_{\0}) = (d \Theta(k)/dk)|_{k = k_{\0}}$ for $t_{\R}(x)$ {\em would} imply a time delay, or a time advance, which depends on the signal of $\Theta^{\prime}(k_{\0})$, for the reflected wave with respect to the incident one, i.e.
\small\begin{equation}
\left.t_{\R}(0) = t_{Inc}(0) -\frac{\Theta^{\prime}(k)}{v_k}\right|_{k = k_{\0}}
\label{p10A}
\end{equation}\normalsize
It is analogous to what happens for the step potential tunneling  penetration when $E < V_{\0}$ \cite{Coh77}.
When $0 < x < L$ in region II, the times for the ``intermediary'' $\alpha$ and $\beta$ wave packet peaks would be given by
\small\begin{equation}
t_{\A}(x) =\left.\left[ \frac{(x - L)}{v_q}
 - \frac{(x_{\0} - \Theta^{\prime}(k))}{v_k}\right]\right|_{k = k_{\0}}~~~~\mbox{and}~~~~
t_{\B}(x)  = \left.\left[ -\frac{(x - L)}{v_q}
-  \frac{(x_{\0} - \Theta^{\prime}(k))}{v_k}\right]\right|_{k = k_o}
\label{p12}
\end{equation}\normalsize
And finally, when $x > L$ in region III, the peak of the transmitted wave packet would be written as
\small\begin{equation}
t_{\T}(x) =\left.\left[ \frac{x - x_{\0} - L + \Theta^{\prime}(k)}{v_k} \right]\right|_{k = k_{\0}}
\label{p14}
\end{equation}\normalsize
The above results (\ref{p9}-\ref{p14}) could be directly obtained by applying the SPM to the wave functions $\psi_f(x, t)$ expressed in Eq.~(\ref{p6A}), where the stationary phases for each decomposed wave component $f$ would be given by
\small\begin{eqnarray}
\vartheta_{Inc}(x,t,k)&=&- E  \, t + k \,(x - x_{\0}),  \nonumber\\
\vartheta_{\R}(x,t,k) &=&- E  \, t - k \,(x + x_{\0}) + \Theta(k),\nonumber\\
\vartheta_{\A}(x,t,k) &=&- E  \, t - k \,x_{\0} + q  (x  - L) + \Theta(k),\nonumber\\
\vartheta_{\B}(x,t,k) &=&- E  \, t - k \,x_{\0}- q  (x - L) + \Theta(k),\nonumber\\
\vartheta_{\T}(x,t,k) &=&- E  \, t + k (x - x_{\0} - L)  + \Theta(k).
\label{p7}
\end{eqnarray}\normalsize
Meanwhile, these time-dependencies must be {\em carefully interpreted}.
In the Fig.(\ref{fig1A}), we illustrate the wave packet diffusion described analytically by the results of Eq.~(\ref{p6C}).
The ``pictures'' display the wave function in the {\em proximity} of the barrier for suitably chosen times.
By taking separately the right ($\alpha$) and left ($\beta$) moving components in region II, independent of the value of $\Theta(k)$, we can observe that
\small\begin{equation}
\Delta t_{\A} =  t_{\A}(L) -  t_{\A}(0) = \Delta t_{\B} =  t_{\B}(0) -  t_{\B}(L)= \left.\frac{L}{v_q} \right|_{k = k_{\0}}
\label{p13}
\end{equation}\normalsize
which should corresponds to transit time values ``classically'' expected.
However, by considering the information carried by the wave packet peaks, we would have a complete time discontinuity at $x = 0$ represented by Eq.~(\ref{p10A}) and by the fact that
\small\begin{equation}
t_{\A}(0) = - \left.\left[\frac{x_{\0}}{v_k} + \frac{L}{v_{q}}
 - \frac{\Theta^{\prime}(k)}{v_k} \right] \right|_{k = k_{\0}} ~~~\neq~~~t_{Inc}(0),
\label{p15}
\end{equation}\normalsize
i. e. the $\alpha$ wave could appear in region II at a time $t_{\A}(0)$ before (or after) the incident wave having arrived at $x = 0$.
Both results expressed by Eqs.~(\ref{p10A}) and (\ref{p15}) which are illustrated in Fig.~\ref{fig1A} should imply the non-conservation of probabilities.
It is, by principle, unacceptable.

\begin{figure}
\begin{center}
\epsfig{file= 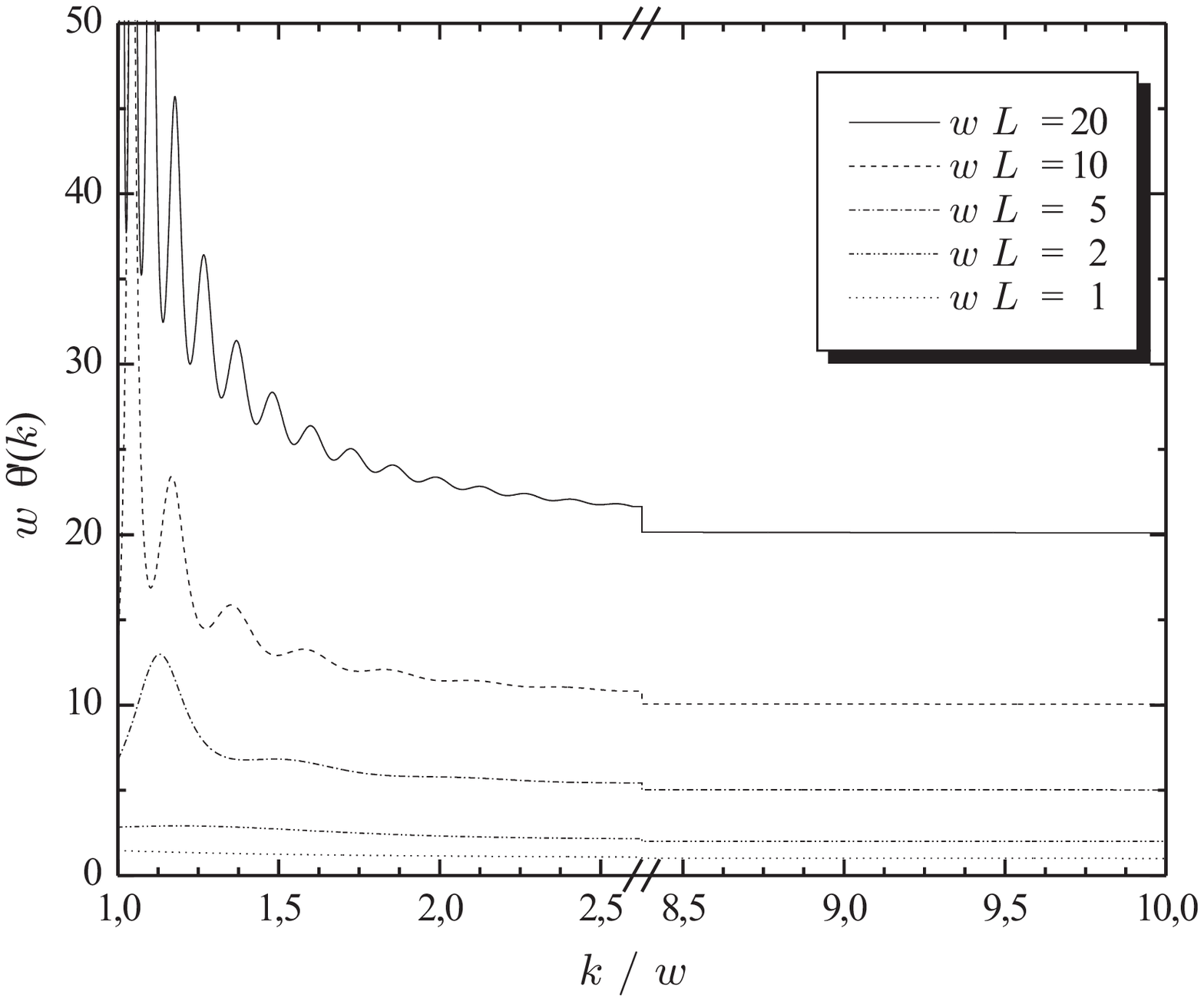, height= 13.5 cm, width = 15 cm}
\end{center}
\caption{\footnotesize
The phase derivative $\Theta^{\prime}(k)$ dependence on $k/w$.
$\Theta^{\prime}(k)$ does not present an adequate analytical behavior (smoothness)
for the applicability of the SPM when $k$ approximates to $w$ since the phase derivative oscillates too rapidly (The phase is not stationary).
The method can be accurately applied for larger values of $k/w$ when the phase is really stationary.}
\label{fig2A}
\end{figure}

\begin{figure}
\begin{center}
\epsfig{file= 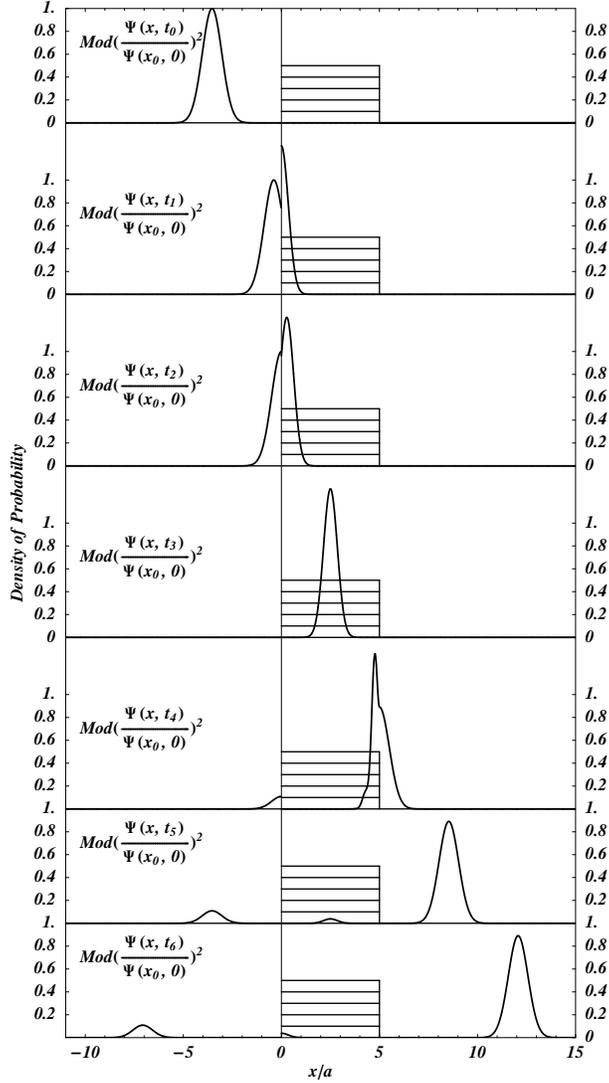, height= 15 cm, width = 8 cm}
\end{center}
\caption{\footnotesize Erroneous interpretation of the scattering of an incoming wave packet by a unidimensional potential
barrier of height $V_{\0}$ and width $L$.
It has the purpose of illustrating the deficiencies inherent to the {\em wrong} applicability of the SPM.
We have plotted the propagating wave packets at the corresponding times $t_{n} = (m a^{\2})[n\,(L/a)]/(a\, q_{\0})$ (with $n = 0,\,1,\,...,\,5$ and with the normalization constraint $m a^{\2} = 1$) by assuming the incoming wave
packet starts at $x =- (k_{\0}\,L)/(2q_{\0})$.
From the {\em false} behavior of the density of probabilities it becomes obvious that the total probability is not conserved as it was expected.
The square of the amplitude modulus would supposedly represent a collision of a wave packet of average width $a$
with a potential barrier $V_{\0}$ of width $L = 5\,a$ where, for illustrating reasons, we have adopted $k_{\0} = \sqrt{2}w$ and $w a = 10000$.
Only a fixed region in $x$ close to the barrier is shown.}
\label{fig1A}
\end{figure}

To be more accurate with this analysis and clear up some dubious points, it should be more convenient to verify two simple particular cases.
In order to simplify the calculations, we set $x_{\0} = 0 $ and we choose $q_{\0} \,  L = n  \pi ~~ (n = 1,2,3,...)$ so that we can write,
\small\begin{equation}
\mathcal{F}(k_{\0}) = 2 k_{\0} q_{\0} , ~~~~\Theta(k_{\0}) = 0,
~~~~ R(k_{\0}) = 0, ~~~~T(k_{\0}) = 1
\label{p16}
\end{equation}\normalsize
and
\small\begin{equation}
\Theta^{\prime}(k) = \frac{m \, L}{q_{\0}}
\frac{k^{\2}_{\0} + q^{\2}_{\0} }{2 k_{\0} q_{\0} } > \frac{m\,  L}{q_{\0} }
\label{p19A}
\end{equation}\normalsize
so that
\small\begin{equation}
t_{\A}(0) = \frac{m\,  L}{q_{\0} }
\frac{\left(k_{\0} - q_{\0} \right)^{\2}}{2 k_{\0} q_{\0} } > 0.
\end{equation}\normalsize
Otherwise, if we choose $q_{\0} \,  L = (n + \frac{1}{2})  \pi  $, we shall obtain,
\small\begin{eqnarray}
&&\mathcal{F}(k_{\0}) = k^{\2}_{\0} + q^{\2}_{\0} ,
~~~~\Theta(k) = \frac{\pi}{2},~~~~
R(k_{\0}) = \frac{k^{\2}_{\0} - q^{\2}_{\0} }{k^{\2}_{\0} + q^{\2}_{\0} },
~~~~T(k_{\0}) = \frac{2 k_{\0} \, q_{\0} }{k^{\2}_{\0} + q^{\2}_{\0} }
\exp{\left[i\left(\frac{\pi}{2} - k_{\0} \, L\right)\right]}
\label{p19C}
\end{eqnarray}\normalsize
and
\small\begin{equation}
\Theta^{\prime}(k) = \frac{m \, L}{q_{\0} }\frac{2 k_{\0} \,  q_{\0} }
{k^{\2}_{\0} + q^{\2}_{\0} } < \frac{m \, L}{q_{\0} }
\label{p19D}
\end{equation}\normalsize
which gives
\small\begin{equation}
t_{\A}(0) = -\frac{m L}{q_{\0} }
\frac{\left(k_{\0} - q_{\0} \right)^{\2}}{k^{\2}_{\0} + q^{\2}_{\0} } < 0.
\end{equation}\normalsize
The latter {\em negative} value corroborates with the absurd possibility of the peak associated with the $\alpha$ wave appears before the peak of the incident wave packet arrival at $x = 0$.
In fact, we must observe that if we proceed with a more general analysis, the phase derivative $\Theta^{\prime}(k)$ does not present a regular behavior.
It oscillates very rapidly, as we can observe in Fig.(\ref{fig2A}) so that the SPM {\em cannot} be applied!

Other incongruities in the naive application of the SPM presented above have been pointed out \cite{Ber04}.
Differently from the analytical analysis presented above, the numerical simulations of a wave packet diffusion above the potential barrier shows the appearance of multiple peaks due to the two reflection points at $x = 0$ and $x = L$ (see the figures).
In fact, numerical calculations automatically conserve probabilities, at least to within the numerical errors.
This observation suggest a new analysis and subsequent interpretation of the ambiguities presented in the previous section.
In order to correctly apply the SPM and accurately obtain the position of the peak of the propagating wave packet with an accurate time dependence, we are constrict to solve the continuity constraints of the Schroedinger equation at each potential discontinuity point $x = 0$ and $x = L$ by considering multiple successive reflections and transmissions.
The phenomenon was already described in \cite{Ber04} and consists in an incoming wave of unitary amplitude with momentum distribution centered at $k_{\0}$ which reaches the interface $x = 0$ where, at an instant $t = t_{\0}$, is decomposed into a reflected wave component of amplitude (see the diagram) $R_{\1}$ and a transmitted wave component of amplitude $\alpha_{\1}$.
The transmitted wave continues propagating until it reaches $x = L$ where, at $t = t_{\1}$, it is now decomposed into a reflected wave component of amplitude $\beta_{\1}$ and a transmitted wave component of amplitude $T_{\1}$.
That new reflected wave component comes back to $x = 0$ where, at $t = t_{\2}$, it is again decomposed into another reflected wave component of amplitude $\alpha_{\2}$ and transmitted wave component of amplitude $R_{\2}$ which will be added to $R_{\1}$.
This iterative process continues for infinity times as we can see in the following diagram,
\small
\[
\begin{array}{l}
\begin{array}{r|l c r|l }
\exp[i\,k\,x] +  R_{\1}  \exp[-i\,k\,x] & \, \alpha_{\1}  \exp[i\,q\,x]&
 & \alpha{\1} \exp[i\,q\,x] \,+\,  \beta{\1}  \exp[-i\,q\,x]  & \, T_{\1} \exp[i\,k\,x]\\
 R_{\2}  \exp[-i\,k\,x] & \, \alpha{\2}  \exp[i\,q\,x] \,+\,  \beta{\1}  \exp[-i\,q\,x]&
 & \hspace*{0.5cm} \alpha{\2} \exp[i\,q\,x] \,+ \, \beta{\2}  \exp[-i\,q\,x]
& \, T_{\2} \exp[i\,k\,x]\\
\vdots \hspace*{1.2cm} &\hspace*{.75cm} \vdots \hspace*{.95cm}+ \hspace*{.7cm}
\vdots \hspace*{1.2cm}&  & \vdots \hspace*{.9cm}+ \hspace*{.7cm}
\vdots \hspace*{1.2cm}& \hspace*{.6cm}\vdots
\\
R_{\n}  \exp[-i\,k\,x] & \, \alpha_{\n} \exp[i\,q\,x] +  \beta_{\n - \1} \exp[-i\,q\,x]&
 & \hspace*{.5cm} \alpha_{\n} \exp[i\,q\,x] +  \beta_{\n}  \exp[-i\,q\,x]
& \, T_{\n} \exp[i\,k\,x]
\end{array}\\
\hspace*{3.35cm} x = 0 \hspace*{8.4cm} x = L
\end{array}
\]
\normalsize
The continuity constraints over $\psi(x,t)$ for each potential step at
$x = 0$ and $x = L$ determine the coefficients
\small\begin{eqnarray}
&& R_{\1} = \frac{k - q}{k + q}\, ,~~~~ \alpha{\1}=
\frac{2\,k}{k + q}\, ,~~~~ \, \, \, \, \beta{\1}= \frac{2\,k (q -
k)}{(k + q)^{\2}}\, \exp[2 \,i\,q\,L\,]\, ,\hspace{5cm}\nonumber\\
&&T_{\1}=
\frac{4 \,k  q}{(k + q)^{\2}}\,\exp[i\,(q - k)\,L] \, ,~~~~
R_{\2} =\frac{q}{k}\, \alpha{\1} \, \beta{\1},
\label{p19BB}
\end{eqnarray}\normalsize
and establish the recurrence relations
\small\begin{eqnarray}
\frac{R_{\n + \2}}{R_{\n + \1}} = \frac{\alpha_{\n + \1}}{\alpha_{\n}} =
\frac{\beta_{\n + \1}}{\beta_{\n}} = \frac{T_{\n + \1}}{T_{\n}} =\left(
\frac{k - q}{k + q}\right)^{\2} \, \exp [2\, i\,q\,L\,] \hspace*{1cm}
\mbox{\small $n=1,2,\dots$}\, \,,\hspace*{2.4cm}
\label{p19B}
\end{eqnarray}\normalsize
which allow us to write the sum of the coefficients
$R_{\n}$, $\alpha_{\n}$,
$\beta_{\n}$, and $T_{\n}$ as
\small\begin{eqnarray}
R      &=&  \sum_{n = \1}^{\infty}R_{\n} =
R_{\1} + R_{\2}  \left[ 1 - \left(
\frac{k - q}{k + q}\right)^{\2} \, \exp [2\, i\,q\,L\,] \right]^{- \1}\,
,\nonumber \\
 \alpha
&=& \sum_{n = \1}^{\infty}\alpha{\n} = \alpha{\1}\left[ 1 - \left( \frac{k
- q}{k + q}\right)^{\2} \, \exp [2\, i\,q\,L\,] \right]^{- \1}\, \, ,\nonumber\\
 \beta
&=& \sum_{n = \1}^{\infty}\beta{\n} = \beta{\1}\left[ 1 - \left( \frac{k
- q}{k + q}\right)^{\2} \, \exp [2\, i\,q\,L\,] \right]^{- \1} \, \, ,\nonumber\\
 T
&=& \sum_{n = \1}^{\infty}T_{\n}\, = \, T_{\1} \left[ 1 - \left(
\frac{k - q}{k + q}\right)^{\2} \, \exp [2\, i\,q\,L\,] \right]^{-
\1}\, \, .
\label{p19}
\end{eqnarray}\normalsize
These sums reproduce exactly the expressions in Eq.~(\ref{p5}).
In this form the interpretation is easy. $R_{\1}$ represents the first reflected wave (it has no time delay since it is real).
$R_{\2}$ represents the second reflected wave and, as a consequence of the continuity condition at $x = 0$, it is the sum, in region II, of the first left-going wave ($\beta{\1}$) and the second right-going amplitude $(\alpha{\2})$, i.e.
\[ R_{\2}= \alpha{\2} + \beta{\1} \equiv \frac{q}{k}\, \alpha{\1} \, \beta{\1}\, \, .\]
This structure is that given by considering two ``step functions'' back-to-back.
Thus at each interface the ``reflected'' and ``transmitted'' waves are instantaneous i. e. without any delay time.

Now we can calculate the time at which each transmitted or reflected wave appears by applying the SPM for each component of the total transmitted $T$ or reflected $R$ coefficients.
It will give us the recurrence relation
\small\begin{equation}
t_{n} = t_{n-1} + \frac{m (x -L)}{q_{\0}}
\label{2p29}
\end{equation}\normalsize
which coincides with the ``classically'' predicted value for the velocity of the particle above the barrier.
Indeed the SPM {\em applied separately} to each term in the above series expansion for $R$ yields delay multiples of $2\left.(\mbox{d}q/\mbox{d}E)\right|_{q = q_{\0}}\,L= 2 (m /q_{\0})\,L$.
This agrees perfectly with the fact that since the peak momentum in region II is $q_{\0}$, the $\alpha$ and $\beta$ waves have group velocities of $q_{\0}/m$ and hence transit times (one way) of
$(m / q_{\0})\,L$.
The first transmitted peak appears (according to this version of the SPM) after a time $(m / q_{\0})\, L$, in perfect agreement with the above interpretation.

Is this compatible with probability conservation? It is because of the following identity
\small\begin{equation}
\sum_{n = \1}^{\infty} \left( |R_{\n}|^{\2} + |T_{\n}|^{\2} \right)
=1\, \, .
\end{equation}\normalsize
This result is by no means obvious since it coexists with the well-known result, from the plane wave analysis,
\small\begin{equation}
|R|^{\2} + |T|^{\2} = |\sum_{n = \1}^{\infty}R_{\n} \,|^{\2} + |
\sum_{n = \1}^{\infty}T_{\n} \,|^{\2}= 1\, \, .
\end{equation}\normalsize
To conclude, we can state that the conditions for applying the SPM depend upon the correct manipulation of the amplitude.
A posteriori this seems obvious but the point is that unless we know the number of separate peaks the SPM is ambiguous.
There is, however, a converse to this question.
If the modulating function is such that two or more wave packets overlap, then we cannot treat them separately without considering all the interference effects.
The above barrier analysis is simply a particular example of this ambiguity.

\begin{figure}
\begin{center}
\epsfig{file= 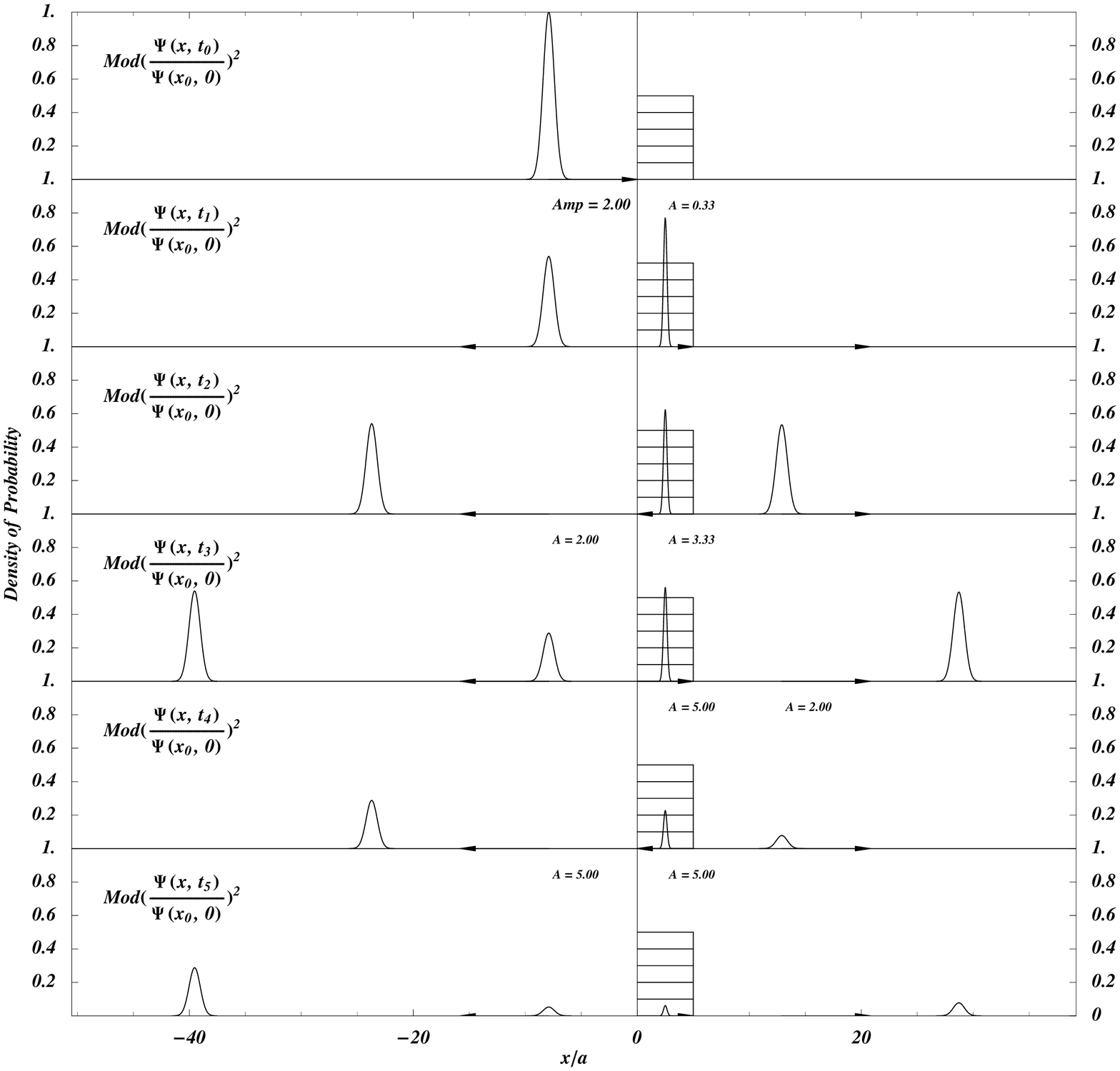, height= 17 cm, width = 15 cm}
\end{center}
\caption{\footnotesize Multiple peak decomposition of a propagating wave packet in the above potential barrier scattering problem.
We have plotted the first few reflected and transmitted waves for
times $t_{n} = (m a^{\2})[n\,(L/a)]/(a\, q_{\0})$ (with $n = 0,\,1,\,...,\,5$ and $m a^{\2} = 1$) in correspondence with Eq.~(\ref{2p29}),
by assuming that the incoming wave packet starts at $x =- (k_{\0}\,L)/(2q_{\0})$.
The density of probabilities represents the collision of a wave packet of average width $a$
with a potential barrier $V_{\0}$ of width $L = 5\,a$.
Just for illustrating reasons, we have adopted $k_{\0} = (\sqrt{10} \,w)/3$ ($w a = 10000$) and we have printed the wave packet
amplification multiplying factor ({\em A}) (individually adopted for visual convenience for each wave packet) when necessary.}
\label{fig3}
\end{figure}

In order to obtain an analytic formula for the wave packet propagation, we consider the {\em gaussian} momentum distribution given by Eq.~(\ref{203}).
Using the linear approximation for the momentum $q$ given by (\ref{p6B}) and considering the expressions in (\ref{p19}) resulting from the multiple peak decomposition in (\ref{p6A}), again, with the same pertinent approximations used for obtaining (\ref{p6C}), we can analytically construct the following simplified expressions
\small\begin{eqnarray}
\psi_{\I}(x,t)&=&\varphi[x - x_{\0},\, t\,],\nonumber\\
\psi_{\R}(x,t) & = & \mbox{$\frac{k_{\0} - q_{\0}}{k_{\0} +
q_{\0}}$}\,\varphi[-\,x-x_{\0},\, t\,]+\nonumber\\
&&
\mbox{$\frac{4k_{\0}q_{\0}(q_{\0}-k_{\0})}{(k_{\0} +
q_{\0})^{\3}}$} \, \exp\left[-i\, \mbox{$\frac{2 w^{\2}}{q_{\0}}$}
\, L\right]\,\sum_{n = \0}^{\infty} \left( \mbox{$\frac{k_{\0} -
q_{\0}}{k_{\0} + q_{\0}}$} \, \exp\left[-i\,
\mbox{$\frac{w^{\2}}{q_{\0}}$} \, L\right] \right)^{\2 n}
\varphi[- x - x_{\0}+2(n+1)\mbox{$\frac{k_{\0}}{q_{\0}}$}\,L,\, t\,],\nonumber\\
\psi_{\A}(x,t) & = & \mbox{$\frac{2k_{\0}}{k_{\0} +
q_{\0}}$}\,\exp\left[-i\, \mbox{$\frac{w^{\2}}{q_{\0}}$} \,
x\right]\,\sum_{n = \0}^{\infty} \left( \mbox{$\frac{k_{\0} -
q_{\0}}{k_{\0} + q_{\0}}$} \, \exp\left[-i\,
\mbox{$\frac{w^{\2}}{q_{\0}}$} \, L\right] \right)^{\2 n}
\varphi[(x+2nL)\,\mbox{$\frac{k_{\0}}{q_{\0}}$} - x_{\0},\, t\,],\nonumber\\
\psi_{\B}(x,t) & = & \mbox{$\frac{2k_{\0}(q_{\0}-k_{\0})}{(k_{\0}
+ q_{\0})^{\2}}$}\,\exp\left[i\, \mbox{$\frac{w^{\2}}{q_{\0}}$}
\, (x-2L)\right]\,\sum_{n = \0}^{\infty} \left(
\mbox{$\frac{k_{\0} - q_{\0}}{k_{\0} + q_{\0}}$} \, \exp\left[-i\,
\mbox{$\frac{w^{\2}}{q_{\0}}$} \, L\right] \right)^{\2 n}
\varphi[(2nL + 2L - x)\,\mbox{$\frac{k_{\0}}{q_{\0}}$} - x_{\0},\, t\,],\nonumber\\
\psi_{\T}(x,t) & = & \mbox{$\frac{4k_{\0}q_{\0}}{(k_{\0} +
q_{\0})^{\2}}$}\, \exp\left[-i\, \mbox{$\frac{w^{\2}}{q_{\0}}$}
\, L\right]\,\sum_{n = \0}^{\infty} \left( \mbox{$\frac{k_{\0} -
q_{\0}}{k_{\0} + q_{\0}}$} \, \exp\left[-i\,
\mbox{$\frac{w^{\2}}{q_{\0}}$} \, L\right] \right)^{\2 n}
\varphi[x - x_{\0}-L+(2n+1)\mbox{$\frac{k_{\0}}{q_{\0}}$}\,L,\, t\,].\nonumber\\
\label{p25}
\end{eqnarray}\normalsize
which allow us to visualize the multiple peak decomposition illustrated in Fig.~\ref{fig3}.
By comparing with the exact numerical results, the analytic approximations expressed in (\ref{p25}) will be valid only under certain restrictive conditions.
To explain such a statement, when we analytically solve the integrals like (\ref{p6A}), the exponential $\exp{[2\, i \, q \, L]}$ which appears in the recurrence relations (\ref{p19B}) are approximated by
\small\begin{equation}
\exp{[2\, i \, q_{\0} \, L + 2\, i \, (k - k_{\0})\frac{k_{\0}}{q_{\0}}]}.
\end{equation}\normalsize
In this case, the above analytic expressions can be obtained only when the exponential function does not oscillate in the interval of relevance of the envelop {\em gaussian} function $g(k - k_{\0})$, i. e. when $\Delta q \, L < \pi$, which can be expressed in terms of the wave packet width as
\small\begin{equation}
\Delta q \, L \, \approx\,  \left.\frac{dq}{dk}\right|_{k = k_{\0}} \, \Delta k \, L \,  \approx\,  \frac{k_{\0}}{q_{\0}} \frac{L}{a} < \pi.
\label{p26}
\end{equation}\normalsize
where we have used $\Delta k \, \approx\, 1/a$.
The above constraint implicitly carries a very peculiar character: the multiple peak decomposition can also be evidenced when the wave packet width $a$ is larger than the potential barrier width $L$, i. e. when we assume the relation $k_{\0}/q_{\0}$ is in the interval $1 < k_{\0}/q_{\0} < \pi$ we can have propagating wave packets with $a > L$ satisfying the requirements for the multiple peak resolution (\ref{p25}).

\section{LIMITATIONS ON THE TUNNELING TIME ANALYSIS}

On the theoretical front, people have tried to introduce quantities that have the dimension
of time and can somehow be associated with the passage of the particle through the barrier or,
strictly speaking, with the definition of the tunneling time.
Since a long time these efforts have led to the introduction of several {\em time} definitions \cite{Olk04,But83,Hau87,Fer90,Yuc92,Hag92,Bro94,Olk95,Jak98,Olk02}, some of which are
completely unrelated to the others, which can be organized into three groups:
(1) The first group comprises a time-dependent description in terms of wave packets where some features
of an incident wave packet and the comparable features of the transmitted packet are
utilized to describe a {\em delay} as tunneling time \cite{Hau89}.
(2) In the second group the tunneling times are computed with basis on averages over a set of kinematical paths,
whose distribution is supposed to describe the particle motion
inside a barrier, i. e. Feynman paths are used like real paths
to calculate an average tunneling time with the weighting
function $\exp{[i\, S\, x(t)/\hbar]}$, where $S$ is the action associated with
the path $x(t)$ (where x(t) represents the Feynman paths initiated
from a point on the left of the barrier and ending at another
point on the right of it \cite{Sok87}).
The Wigner distribution paths \cite{Bro94}, and the Bohm approach \cite{Ima97,Abo00} are included in this group.
(3) In the third group we notice the introduction of a new degree of freedom,
constituting a physical clock for the measurements of tunneling times.
Separately, it stands by itself the dwell time approach.
The time related to the tunneling process is defined by the interval during which the incident flux has to exist and act,
to provide the expected accumulated particle storage, inside the barrier \cite{Lan94}.
The methods with a Larmor clock \cite{But83} or an oscillating barrier \cite{But82} are comprised by this group.

There is no general agreement \cite{Olk04,Olk92} among the above definitions about the meaning of tunneling times
(some of the proposed tunneling times are actually traversal times, while others seem to represent in reality only the
spread of their distributions) and about which, if any, of them
is the proper tunneling time, in particular, due to the following reasons \cite{Olk04}:
(a) the problem of defining tunneling times is closely connected with the more general definition of the
quantum-collision duration, and therefore with the fundamental fact that
in quantum mechanics, time enters as a parameter rather than an observable to which an operator can be assigned
(b) the motion of particles inside a potential barrier is a quantum phenomenon, that till now has been devoid of any
direct classical limit;
(c) there are essential differences among the initial, boundary and external
conditions assumed within the various definitions proposed in the literature; those differences have
not been sufficiently analyzed yet.
In particular, the study of tunneling mechanisms is embedded by theoretical constructions involving analytically-continuous {\em gaussian}, or infinite-bandwidth step pulses to examine the tunneling process.
Nevertheless, such holomorphic functions do not have a well-defined front in a manner that the interpretation of the wave packet speed of propagation becomes ambiguous.
Moreover, infinite bandwidth signals cannot propagate through any real physical medium (whose transfer function is
therefore finite) without pulse distortion, which also leads to ambiguities in determining the propagation velocity during the tunneling process.
For instance, some of the barrier traversal time definitions lead, in tunneling time conditions, to very short times which can even become negative, precipitately
inducing to an interpretation of violation of simple concepts of causality.
Otherwise, negative speeds do not seem to create problems with causality,
since they were predicted both within special relativity and within quantum mechanics \cite{Olk95}.
A possible explanation of the time advancements related to the negative speeds can come, in any case, from consideration of the very rapid
spreading of the initial and transmitted wave packets for large momentum distribution widths.
Due to the similarities between tunneling (quantum) packets and evanescent (classical) waves,
exactly the same phenomena are to be expected in the case of classical barriers
(we can mention the analogy between
the stationary Helmholtz equation for an electromagnetic wave packet - in a waveguide, for instance - in
the presence of a {\em classical} barrier and the stationary Schroedinger equation, in the presence of a potential barrier \cite{Lan94,Jak98,Nim94}).
The existence of such negative times is predicted by relativity itself, on the basis of the ordinary postulates \cite{Olk04}, and they appear to have been experimentally detected in many
works \cite{Gar70,Chu82}.

In this extensively explored scenario,
the first group quoted above contains the so-called phase times \cite{Boh52,Wig55}
which are obtained when the stationary phase method (SPM)
is employed for obtaining the times related to the motion of the wave packet spatial centroid
which adopts averages over fluxes pointing in a well-defined direction only, and has recourse to a quantum operator
for time \cite{Olk04}.
Generically speaking, the SPM essentially enables us
to parameterize some subtleties of several quantum phenomena,
such as tunneling \cite{Hau89,Ste93,Bro94},
resonances \cite{Bra70}, incidence-reflection
and incidence-transmission interferences \cite{Per01}
as well as the Hartman Effect \cite{Har62} and
its {\em superluminal} traversal time interpretation \cite{Olk04,Lan94,Jak98}.
In fact, it is the simplest and commonest approximation method for describing the group velocity of
a wave packet in a quantum scattering process represented by a collision of a particle with a potential barrier
\cite{Olk04,Lan94,Hau87,Wig55,Har62,Ber04}.

In the following, our attention is particularly concentrated on some limitations on the use of the SPM for deriving tunneling times
for which we furnish an accurate quantification of the analytical incongruities which restrict the applicability of this method.
We introduce a theoretical construction involving a symmetrical collision with a unidimensional square potential where the
scattered wave packets can be reconstructed by summing the amplitudes of the reflected and transmitted waves
in the scope of what we denominate a multiple peak decomposition analysis \cite{Ber04} in a manner that
the analytical conditions for the SPM applicability are totally recovered.

Generically speaking, the SPM can be successfully utilized for describing
the movement of the center of a wave packet constructed in terms of a momentum distribution $g(k - k_{\0})$
which has a pronounced peak around $k_{\0}$.
By assuming the phase which characterizes the propagation varies sufficiently smoothly
around the maximum of $g(k - k_{\0})$, the stationary phase condition enable us to calculate the position of the
peak of the wave packet (highest probability region to find the propagating particle).
With regard to the tunneling effect, the method is indiscriminately applied to find the position of a wave packet
which traverse a potential barrier.

For the case we consider the potential barrier
\small\begin{equation}
V(x) = \left\{\begin{array}{cll} V_o && ~~~~x \in \mbox{$\left[- L/2, \, L/2\right]$}\\ &&\\ 0&& ~~~~x \in\hspace{-0.3cm}\slash\hspace{0.1cm}\mbox{$\left[- L/2, \, L/2\right]$}\end{array}\right.
\label{2p60}
\end{equation}\normalsize
it is well known that the transmitted wave packet solution ($x \geq L/2 $) calculated by means of the Schroedinger formalism is
given by \cite{Coh77}
\small\begin{equation}
\psi^{\T}(x,t) = \int_{_{\0}}^{^{w}}\frac{dk}{2\pi} \, g(k - k_{\0}) \, |T(k, L)|\,
\exp{\left[ i \, k \,(x - L/2) - i \, \frac{k^2}{2\,m} \, t +
 i \,\Theta(k, L)\right]}
\end{equation}\normalsize
where, in case of tunneling, the transmitted amplitude is written as
\small\begin{equation}
|T(k, L)| =
\left\{1+ \frac{w^4}{4 \, k^2 \, \rho^{\2}(k)}
\sinh^2{\left[\rho(k)\, L \right]}\right\}^{-\frac{1}{2}},
\label{1}
\end{equation}\normalsize
and the phase shift is described in terms of
\small\begin{equation}
\Theta(k, L) = \arctan{\left\{\frac{2\, k^2 - w^2}
{k \, \rho(k)}
\tanh{\left[\rho(k) \, L \right]}\right\}},
\label{502}
\end{equation}\normalsize
for which we have made explicit the dependence on the barrier length $L$ and
we have adopted $\rho(k) = \left(w^2 - k^2\right)^{\frac{1}{2}}$ with $w = \left(2\, m \,V_{\0}\right)^{\frac{1}{2}}$ and $\hbar = 1$.
Without thinking over an eventual distortion that $|T(k, L)|$ causes to the supposedly symmetric function $g(k - k_{\0})$,
when the stationary phase condition is applied to the phase of Eq.~(\ref{1}) we obtain
\small\begin{eqnarray}
\frac{d}{dk}\left.\left\{k \,(x - L/2) - \frac{k^2}{2\,m} \, t
+ \Theta(k, L)\right\}\right|_{_{k = k_{\mbox{\tiny max}}}}
&=&0 ~~~~\Rightarrow\nonumber\\
  x - L/2 - \frac{k_{\mbox{\tiny max}}}{m} \, t +
\left.\frac{d\Theta(k, L)}{dk}\right|_{_{k = k_{\mbox{\tiny max}}}} &=& 0.
\label{3}
\end{eqnarray}\normalsize
The above result is frequently adopted for calculating the transit time $t_{T}$ of
a transmitted wave packet when its peak emerges at $x = L/2$,
\small\begin{equation}
t_{T} =\frac{m}{k_{\mbox{\tiny max}}}\left.\frac{d\Theta(k, \alpha_{(\L)})}{dk}\right|_{_{k = k_{\mbox{\tiny max}}}} =
\frac{2\,m \, L}{k_{\mbox{\tiny max}} \,\alpha }
\left\{\frac{w^4\,\sinh{(\alpha)}\cosh{(\alpha)}
-\left(2\, k_{\mbox{\tiny max}}^2 - w^2 \right)k_{\mbox{\tiny max}}^2 \,\alpha }
{4\, k_{\mbox{\tiny max}}^2 \,\left(w^2 - k_{\mbox{\tiny max}}^2 \right)  +
w^4\,\sinh^2{(\alpha)}}\right\}
\label{4}
\end{equation}\normalsize
where we have defined the parameter
$\alpha = \left(w^2 - k_{\mbox{\tiny max}}^2 \right)^{\frac{1}{2}}\, L$.
The concept of {\em opaque} limit is introduced when
we assume that $k_{\mbox{\tiny max}}$ is independent of $L$
and then we make $\alpha$ tends to $\infty$ \cite{Jak98}.
In this case, the transit time can be rewritten as
\small\begin{equation}
t^{^{OL}}_T = \frac{2\,m}{k_{\mbox{\tiny max}}\,\rho(k_{\mbox{\tiny max}})}.
\label{5}
\end{equation}\normalsize
In the literature, the value of $k_{\mbox{\tiny max}}$ is frequently
approximated by $k_{\0}$, the maximum of $g(k - k_{\0})$, which, in fact,
does not depend on $L$ and could lead us
to the transmission time {\em superluminal} interpretation \cite{Jak98,Olk92,Esp03} of (\ref{5}).
To clear up this point, we notice that when we take the {\em opaque} limit with $L$ going to $\infty$ and
$w$ fixed
as well as with $w$ going to $\infty$ and $L$ fixed, with $k_{\0} < w$ in
both cases, the expression (\ref{5})
suggests times corresponding to a transmission process performed with velocities higher than $c$ \cite{Jak98}.
Such a {\em superluminal} interpretation was extended to the study
of quantum tunneling through
two successive barriers separated by a free region where it was theoretically
verified that the total traversal time does not depend not only on the barrier
widths, but also on the free region extension \cite{Olk92}.
Besides, in a subsequent analysis,
the same technique was applied to a problem with multiple successive barrier
where the tunneling process was designated as a highly non-local phenomenon
\cite{Esp03}.

It would be perfectly acceptable to consider $k_{\mbox{\tiny max}} = k_{\0}$
for the application of the stationary phase condition
when the momentum distribution $g(k - k_{\0})$ centered at $k_{\0}$ is not
modified by any bound condition. This is the case of the incident wave packet
before colliding with the potential barrier.
Our criticism concern, however, with the way of obtaining all the above results
for the transmitted wave packet.
It does not have taken into account the bounds and enhancements imposed
by the analytical form of the transmission coefficient.

Hartman himself had already noticed that the

\vspace{0.3cm}
{\it`` Discussion of the time dependent behavior of a wave packet
is complicated by its diffuse or spreading nature;
however, the position of the peak of a {\em symmetrical} packet can
be described with some precision''}
\vspace{0.3cm}

To perform the correct analysis, we should calculate the right value of
$k_{\mbox{\tiny max}}$ to be substituted in Eq.~(\ref{4}) before taking
the {\em opaque} limit.
We are obliged to consider the relevant amplitude for the transmitted wave as
the product of a symmetric momentum distribution $g(k - k_{\0})$ which describes the
{\em incoming} wave packet by the modulus of the transmission amplitude
$T(k, L)$ which is a crescent function of $k$.
The maximum of this product representing the transmission
modulating function would be given by the solution of the equation
\small\begin{eqnarray}
g(k - k_{\0}) \,\left|T(k, L)\right|\,\left[\frac{g^{\prime}(k - k_{\0})}{g(k - k_{\0})}+
\frac{\left|T(k, L)\right|^{\prime}}{\left|T(k, L)\right|}\right] &=& 0
\label{3p42B}
\end{eqnarray}\normalsize
Obviously, the peak of the
modified momentum distribution is shifted to the right of $k_{\0}$
so that $k_{\mbox{\tiny max}}$ have to be found in the interval $]k_{\0}, w[$.
Moreover, we confirm that $k_{\mbox{\tiny max}}$ presents an implicit
dependence on $L$ as we can demonstrate by the numerical results
presented in {\it Table 1} where we have found the maximum of
$g(k - k_{\0}) \, |T(k, L)|$ by assuming $g(k - k_{\0})$ is a {\em gaussian} function
almost completely comprised in the interval $[0, w]$ given by
\small\begin{equation}
g(k - k_{\0}) = \left(\frac{a^2}{2 \, \pi}\right)^{\frac{1}{4}}
\exp{\left[-\frac{a^2 (k -k_{\0})^2}{4}\right]}.
\label{6}
\end{equation}\normalsize
By increasing the value of $L$ with respect to $a$,
the value of $k_{\mbox{\tiny max}}$ obtained from the numerical calculations
to be substituted in Eq.~(\ref{4}) also increases until $L$
reaches certain values for which
the modified momentum distribution
becomes unavoidably distorted.
In this case, the relevant values of $k$ are
concentrated around the upper bound value $w$.
We shall show in the following that the value of $L$ which starts
to distort the momentum distribution can be analytically obtained
in terms of $a$.

\begin{table}[ht]
\begin{minipage}{14cm}
\begin{center}
\caption{The values of $k$ numerically obtained
by increasing the barrier extension $L$.The values are calculated in terms of the wave packet width $a$ for different values of the
potential barrier high expressed in terms of $w \, a$.
We have fixed the incoming momentum by setting $k_{\0} \, a$ equal to 1.
}
\begin{tabular}{c|ccccccc}
\hline
\hline
~~~~~~~~~$w \, a$& 1.5 & 2.0 & 4.0 & 6.0 & 8.0 & 10 & 20 \\
$ L / a $ &&&&&&&\\
\hline\hline
0.00 & 1.0000& 1.0000& 1.0000& 1.0000& 1.0000& 1.0000& 1.0000\\
0.05 & 1.0062& 1.0188& 1.1777& 1.4156& 1.6238& 1.7726& 1.9834\\
0.10 & 1.0235& 1.0648& 1.3799& 1.6769& 1.8547& 1.9397& 2.0051\\
0.15 & 1.0489& 1.1223& 1.5349& 1.8251& 1.9505& 1.9937& 2.0133\\
0.20 & 1.0794& 1.1825& 1.6571& 1.9178& 2.0000& 2.0204& 2.0203\\
0.25 & 1.1129& 1.2420& 1.7575& 1.9813& 2.0317& 2.0390& 2.0272\\
0.30 & 1.1478& 1.3001& 1.8430& 2.0289& 2.0562& 2.0551& 2.0342\\
0.35 & 1.1836& 1.3565& 1.9185& 2.0679& 2.0779& 2.0704& 2.0413\\
0.40 & 1.2196& 1.4116& 1.9874& 2.1025& 2.0986& 2.0857& 2.0484\\
0.45 & 1.2558& 1.4657& 2.0524& 2.1350& 2.1191& 2.1012& 2.0556\\
0.50 & 1.2921& 1.5194& 2.1155& 2.1668& 2.1399& 2.1170& 2.0628\\
0.55 & 1.3285& 1.5729& 2.1785& 2.1988& 2.1611& 2.1331& 2.0701\\
0.60 & 1.3649& 1.6266& 2.2429& 2.2314& 2.1828& 2.1495& 2.0775\\
0.65 & 1.4015& 1.6809& 2.3101& 2.2651& 2.2051& 2.1663& 2.0850\\
0.70 & 1.4383& 1.7360& 2.3819& 2.3002& 2.2281& 2.1834& 2.0925\\
0.75 & 1.4751& 1.7920& 2.4599& 2.3367& 2.2518& 2.2009& 2.1001\\
0.80 &    $*\footnote{
For the values of $L$ marked with $*$, we can demonstrate by means of Eqs.~(\ref{b}-\ref{c})
that the
modulated momentum distribution
has already been completely distorted and
the maximum loses its meaning
in the context of applicability of the method of stationary phase.
}
$& 1.8489& 2.5466& 2.3751& 2.2761& 2.2188& 2.1078\\
0.85 &    $*$& 1.9065& 2.6456& 2.4154& 2.3013& 2.2371& 2.1155\\
0.90 &    $*$& 1.9646& 2.7627& 2.4578& 2.3272& 2.2558& 2.1234\\
0.95 &    $*$&    $*$& 2.9091& 2.5028& 2.3540& 2.2750& 2.1313\\
1.00 &    $*$&    $*$& 3.1137& 2.5504& 2.3818& 2.2947& 2.1392\\
\hline\hline
\end{tabular}
\end{center}
\end{minipage}
\end{table}

Now, if we take the {\em opaque} limit of $\alpha$ by fixing $L$
and increasing $w$, the above results immediately ruin the
{\em superluminal} interpretation upon the result of Eq.~(\ref{4})
since $t^{^{OL}}_T$ tends to $\infty$ when $k_{\mbox{\tiny max}}$ is
substituted by $w$.

Otherwise, when $w$ is fixed and $L$ tends to $\infty$,
the parameter $\alpha$ calculated at $k = w$ becomes
indeterminate.
The transit time
$t_T$ still tends to $\infty$ but now it exhibits a peculiar
dependence on $L$ which can be easily observed by defining
the auxiliary function
\small\begin{equation}
G(\alpha) =
\frac{\sinh{(\alpha)}\cosh{(\alpha)} - \alpha}
{\sinh^2{(\alpha)}}
\label{11}
\end{equation}\normalsize
which allow us to write
\small\begin{equation}
t^{\alpha}_T = \frac{2\, m\, L}{w \, \alpha}\,
G(\alpha).
\label{12}
\end{equation}\normalsize
When $\alpha \gg 1$,
the transmission time always assume infinite values
with an asymptotic dependence on $\left(w^2 - k^2 \right)^{-\frac{1}{2}}$,
\small\begin{equation}
t^{\alpha}_T \approx  \frac{2\, m }{w \, \left(w^2 - k^2 \right)^{\frac{1}{2}}} \rightarrow \infty.
\label{13}
\end{equation}\normalsize
Only when $\alpha$ tends to $0$ we have an explicit
linear dependence on $L$ given by
\small\begin{equation}
t^{0}_T = \frac{2\, m \, L}{w}\,\lim_{\alpha \rightarrow 0}
{\left\{\frac{G(\alpha)}{\alpha}\right\}}
= \frac{4\, m \, L}{3 \, w}
\label{14}
\end{equation}\normalsize

In addition to the above results, the transmitted wave must be carefully
studied in terms of the rapport between the barrier extension $L$ and the
wave packet width $a$.
For very thin barriers, i. e. when $L$ is much
smaller than $a$,
the modified transmitted wave packet presents substantially the same form
of the incident one.
For thicker barriers, but yet with $L < a$, the peak of the {\em gaussian}
wave packet modulated by the transmission coefficient is shifted
to higher energy values, i. e.  $k_{\mbox{\tiny max}} > k_{\0}$ increases
with $L$.
For very thick barriers, i. e. when $L > a$, we are able to observe that
the form of the transmitted wave packet
is badly distorted with the greatest contribution coming from
the Fourier components corresponding to the energy $w$
just above the top of the barrier in a kind of
{\em filter effect}.
We observe that the quoted distortion
starts to appear when the modulated momentum distribution presents a {\em local maximal} point at $k = w$
which occurs when
\small\begin{equation}
\left.\frac{d}{dk}\left[g(k - k_{\0}) \,
\left|T(k, L)\right|\right]\right|_{k = w} > 0
\label{a}.
\end{equation}\normalsize
Since the derivative of the {\em gaussian} function $g(k - k_{\0})$ is negative at $k = w$, the Eq.~(\ref{a})
gives the relation
\small\begin{equation}
- \frac{g^{\prime}(w - k_{\0})}{g(w - k_{\0})} < \lim_{k \rightarrow w}
{\left[\frac{T^{\prime}(k, L)}{T(k, L)}\right]}
= \frac{w \, L^2}{4}\frac{\left(1 + \frac{w \, L^2}{3}\right)}
{\left(1 + \frac{w \, L^2}{4}\right)} < \frac{w \, L^2}{3}
\label{b}
\end{equation}\normalsize
which effectively represents the inequality
\small\begin{equation}
\frac{a^2}{2}\,(w - k_{\0}) < \frac{w \,L^2}{3}~~\Rightarrow~~
L > \sqrt{\frac{3}{2}}\,a\, \left(1 - \frac{k_{\0}}{w}\right).
\label{c}
\end{equation}\normalsize

\begin{figure}
\centerline{\psfig{file=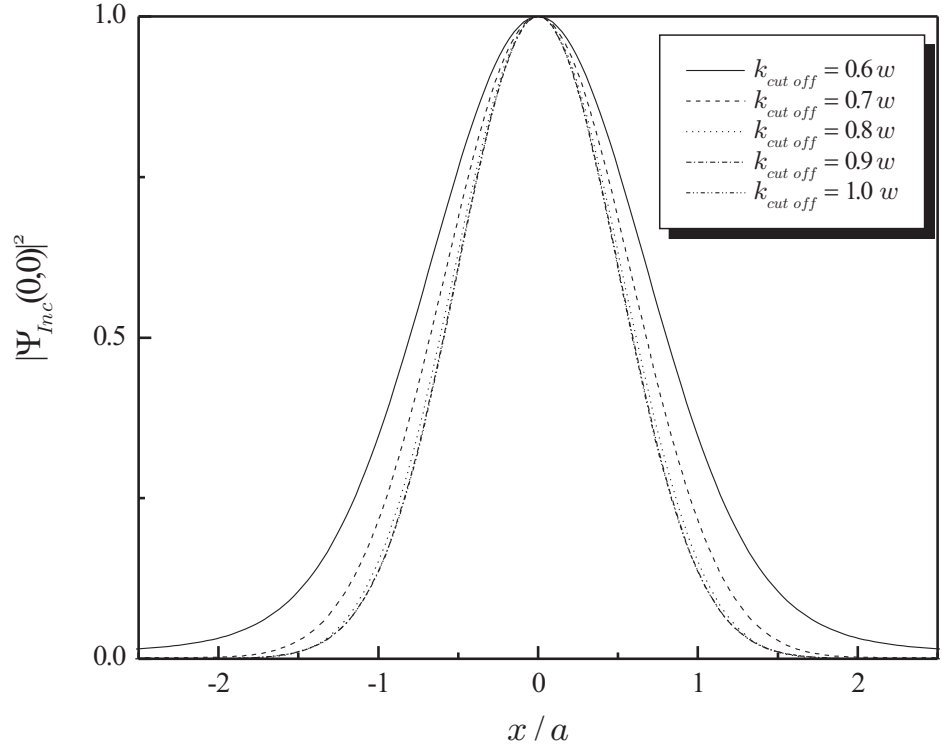,width=14cm}}
\vspace*{8pt}
\caption{\footnotesize Dependence of the wave packet shape on the {\em cut off} value
of a momentum distribution centered around $k_{o} = 0.5 w$ with the values of $k$
comprised between $0$ and $k_{\mbox{\tiny \em{cut off}}}$.\label{fig2}}
\end{figure}

Due to the {\em filter effect}, the amplitude of the transmitted wave
is essentially composed by the plane wave components of the front tail of the
{\em incoming} wave packet which reaches the first barrier interface before
the peak arrival.
Meanwhile, only whether we had {\em cut} the momentum distribution {\em off} at a
value of $k$ smaller than $w$, i. e. $k \approx (1 - \delta) w$,
the {\em superluminal} interpretation
of the transition time (\ref{5}) could be recovered.
In this case, independently of the way as $\alpha$ tends to
$\infty$, the value assumed by the transit time would be approximated by
$t^{\alpha}_{T} \approx 2 \,m / w \, \delta$ which is a finite quantity.
Such a finite value would confirm the hypothesis of {\em superluminality}.
However, the {\em cut off} at $k \approx (1 - \delta) w$ increases the amplitude of
the tail of the incident wave as we can observe in Fig.~\ref{fig2}.
It contributes so relevantly as the peak of the incident wave to the final
composition of the transmitted wave
and creates an ambiguity in the definition of the
{\em arrival} time.

\section{ONE DIMENSIONAL POTENTIAL SYMMETRICAL SCATTERING FOR $E < V_{\0}$}

In order to recover the scattered momentum distribution symmetry conditions for accurately applying the SPM, we
assume a symmetrical colliding configuration of two wave packets traveling in opposite directions.
By considering the same barrier represented in (\ref{2p60}), we solve the Schroedinger equation for
a plane wave component of momentum $k$ for two identical wave packets symmetrically separated from the origin $x = 0$, at time
of collision $t = - (m L) /(2 k_{\0})$ chosen for mathematical convenience, where we assume they perform a totally symmetric simultaneous collision with the potential barrier.
The wave packet reaching the left(right)-side of the barrier is represented by
\small\begin{equation}
\psi^{\L(\R)}(x,t) = \int_{_{\mi\infty}}^{^{\pl\infty}}dk \, g(k - k_{\0})\phi^{\L(\R)}(k,x)\, \exp{[- i \, E\,t]}
\end{equation}\normalsize
where we assume the integral can be naturally extended from the interval $[0,w]$ to the interval $[-\infty,\pl\infty]$ as a first approximation.
By assuming that $\phi^{\L(\R)}(k,x)$ are Schroedinger equation solutions, at the time $t = - (m L) /(2 k_{\0})$ i. e. when the wave packet peaks simultaneously
reach the barrier, we can write
\small\begin{equation}
\phi^{\L(\R)}(k,x)=\left\{
\begin{array}{l l l l}
\phi^{\L(\R)}_{\1}(k,x) &=& \exp{\left[ \pm i \,k \,x\right]} + R^{\L(\R)}_{\bbb}(k,L)\exp{\left[ \mp i \,k \,x\right]}&~~~~x < - L/2\, (x > L/2),\nonumber\\
\phi^{\L(\R)}_{\2}(k,x) &=& \alpha^{\L(\R)}_{\bbb}(k)\exp{\left[ \mp\rho  \,x\right]} + \beta^{\L(\R)}_{\bbb}(k)\exp{\left[ \pm\rho  \,x\right]}&~~~~- L/2 < x < L/2,\nonumber\\
\phi^{\L(\R)}_{\3}(k,x) &=& T^{\L(\R)}_{\bbb}(k,L)\exp{ \left[\pm i \,k \,x\right]}&~~~~x > L/2 \, (x < - L/2) .
\end{array}\right.
\label{510}
\end{equation}\normalsize
where the upper(lower) sign is related to the index $L$($R$).
By assuming the constraints which require the continuity of $\phi^{\L,\R}$ and their derivatives
at $x = - L/2$ and $x = L/2$, after some mathematical manipulations, we can easily obtain
\small\begin{equation}
R^{\L,\R}_{\bbb}(k,L) = \exp{\left[ - i \,k \,L \right]} \left\{\frac{\exp{\left[ i \, \Theta(k,L)\right]} \left[1 - \exp{\left[ 2\,\rho(k) \,L\right]}\right]}{1 - \exp{\left[ 2\,\rho(k) \,L\right]}\exp{\left[ i\, \Theta(k,L)\right]}}\right\}
\label{511}
\end{equation}\normalsize
and
\small\begin{equation}
T^{\L,\R}_{\bbb}(k,L) = \exp{\left[ - i \,k \,L \right]} \left\{\frac{\exp{\left[\rho(k) \,L\right]}\left[1- \exp{\left[ 2\, i\, \Theta(k,L)\right]}\right]}{1 - \exp{\left[ 2\,\rho(k) \,L\right]}\exp{\left[ i\, \Theta(k,L)\right]}}\right\},
\label{512}
\end{equation}\normalsize
where $\Theta(k,L)$ is given by the Eq.~(\ref{502}) and $R^{\L}_{\bbb}(k,L)$ and $T^{\R}_{\bbb}(k,L)$ as well as $R^{\R}_{\bbb}(k,L)$ and $T^{\L}_{\bbb}(k,L)$
are intersecting each other.
By analogy with the amplitude addition procedure we have adopted in the multiple peak decomposition scattering \cite{Ber04},
such a pictorial configuration obliges us to sum the intersecting amplitude of probabilities before taking their squared
modulus in order to obtain
\small\begin{equation}
R^{\L,\R}_{\bbb}(k,L)+ T^{\R,\L}_{\bbb}(k,L) = \exp{\left[ - i \,k \,L \right]} \left\{\frac{\exp{\left[\rho(k) \,L\right]}+ \exp{\left[ i\, \Theta(k,L)\right]}}{1 + \exp{\left[ \rho(k) \,L\right]}\exp{\left[ i\, \Theta(k,L)\right]}}\right\}
 = \exp{\left\{ - i [k \,L - \varphi(k,L)]\right\}}
\label{513}
\end{equation}\normalsize
with
\small\begin{equation}
\varphi(k,L) = -\arctan{\left\{\frac{2\,k\,\rho(k) \, \sinh{[\rho(k)\,L]}}{w^{\2} + \left(k^{\2}-\rho^{\2}(k)\right)\cosh{[\rho(k)\,L]}}\right\}}.
\label{514}
\end{equation}\normalsize
The important information we get from the relation given by (\ref{513}) is that,
differently from the previous standard tunneling analysis, by adding the intersecting amplitudes at each side of the barrier,
we keep the original momentum distribution undistorted since the modulus $|R^{\L,\R}_{\bbb}(k,L)+ T^{\R,\L}_{\bbb}(k,L)|$ is unitary.
At this point we recover the most fundamental condition for the applicability of the SPM which allows us to
accurately find the position of the peak of the reconstructed wave packet composed by reflected and transmitted
superposing components.

The phase time interpretation can be, in this case, correctly quantified in terms of the analysis of the
{\em new} phase $\varphi(k, L)$.
By applying the stationary phase condition to the recomposed wave packets, the maximal point of the
scattered amplitudes $g(k - k_{\0})|R^{\L,\R}_{\bbb}(k,L)+ T^{\R,\L}_{\bbb}(k,L)|$ are accurately given by $k_{\mbox{\tiny max}} = k_{\0}$
so that the traversal/reflection time or, more generically, the scattering time, results in
\small\begin{equation}
t^{^{\varphi}}_{T} =\frac{m }{k_{\0}}\left.\frac{d\varphi(k, \alpha_{(\L)})}{dk}\right|_{_{k = k_{\0}}} =
\frac{2\,m\, L}{k_{\0}\,\alpha}
\frac{w^{\2}\sinh{(\alpha)} + \alpha\,k^{\2}_{\0}}{2\,k^{\2}_{\0} - w^{\2} + w^{\2}\cosh{(\alpha)}}
\label{515}
\end{equation}\normalsize
with $\alpha$ previously defined.
It can be said metaphorically that the identical particles represented by both incident wave packets spend a time of the order of $ t^{^{\varphi}}_{T}$
inside the barrier before retracing its steps or tunneling.
In fact, we cannot differentiate the tunneling from the reflecting waves for such a scattering configuration.
The point is that we have introduced a possibility of improving
the efficiency of the SPM in calculating reflecting and tunneling phase times
by studying a process where the conditions for applying the method are totally recovered,
i. e. we have demonstrated that the transmitted and reflected interfering amplitudes results
in a unimodular function which just modifies the {\em envelop} function $g(k - k_{\0})$ by an additional phase.
The previously observed incongruities which cause the distortion of the momentum distribution $g(k - k_{\0})$ are
completely eliminated in this case.
At the same time, one could argue about the possibility of extending such a result to the standardly established
tunneling process.
We should assume that in the region inside the potential barrier, the reflecting and transmitting
amplitudes should be summed before we compute the phase changes.
Obviously, it would result in the same phase time expression represented by (\ref{515}).
In this case, the assumption of there (not) existing interference between the momentum amplitudes of the reflected and transmitted
at the discontinuity points $x = -L/2$ and $x = L/2$ is purely arbitrary.
Consequently, it is important to reinforce the argument that such a possibility of interference leading to different phase time results
is strictly related to the idea of using (or not) the multiple peak (de)composition in the region where the potential barrier is localized.

\begin{figure}
\centerline{\psfig{file=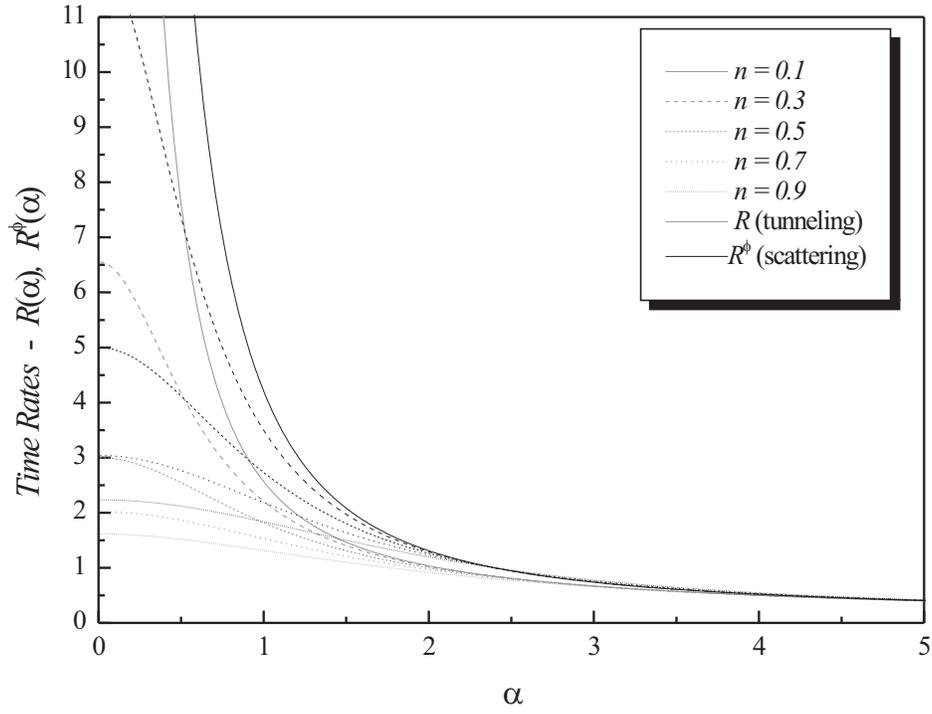,width=14cm}}
\vspace*{8pt}
\caption{\footnotesize Time rates for the {\em standard} tunneling and the {\em new}
scattering process.
The rates $R(\alpha)$ and $R^{\phi}(\alpha)$ can be understood as transmitted times in the unities of the
classical propagation time $\tau$. Both of them present the same asymptotic behavior.
\label{fig3A}}
\end{figure}

In order to illustrate the difference between the standard
{\em tunneling} phase time $t_{T}$ and the alternative
{\em scattering} phase time $t^{^{\varphi}}_{T}$ we introduce the new parameter $n = k^{\2}_{\mbox{\tiny max}}/w^{\2}$
and the {\em classical} traversal time $\tau = (m L) /k_{\mbox{\tiny max}}$ in order to define the rates
\small\begin{equation}
R_{T}(\alpha) = \frac{t_{T}}{\tau}=
\frac{2}{\alpha}\left\{
\frac{\cosh{(\alpha)}\sinh{(\alpha)} - \alpha\,n\left(2 n - 1\right)}{\left[4 n \left(1 - n\right)+\sinh^{\2}{(\alpha)}\right]}
\right\}\label{516}
\end{equation}\normalsize
and
\small\begin{equation}
R^{^{\varphi}}_{T}(\alpha) = \frac{t^{^{\varphi}}_{T}}{\tau}=
\frac{2}{\alpha}\left\{\frac{n\, \alpha + \sinh{(\alpha)}}{2n - 1 +\cosh{(\alpha)}}
\right\}\label{517}
\end{equation}\normalsize
which are plotted in the Fig.(\ref{fig3A}) for some discrete values of $n$ varying from $0.1$ to $0.9$,
from which we can obtain the most common limits given by
\small\begin{equation}
\lim_{\alpha \rightarrow \infty}
{\left\{R^{^{\varphi}}_{T}(\alpha)\right\}}
=
\lim_{\alpha \rightarrow \infty}
{\left\{R_{T}(\alpha)\right\}}
=0
\label{518}
\end{equation}\normalsize
and
\small\begin{equation}
\lim_{\alpha \rightarrow 0}
{\left\{R_{T}(\alpha)\right\}}
= 1+ \frac{1}{2 n}, ~~~~
\lim_{\alpha \rightarrow 0}
{\left\{R^{^{\varphi}}_{T}(\alpha)\right\}}
= 1+ \frac{1}{n}
\label{519}
\end{equation}\normalsize

Both of them present the same asymptotic behavior which, at first glance, recover the theoretical possibility of a {\em superluminal} transmission in the sense that, by now, the SPM can be correctly applied since the analytical limitations are accurately observed.
At this point, it is convenient to notice that the superluminal phenomena, observed in the experiments with tunneling photons and evanescent electromagnetic waves \cite{Nim92,Ste93,Chi98,Hay01}, has generated a lot of discussions on relativistic causality.
In fact, superluminal group velocities in connection with quantum (and classical) tunnelings were predicted even on the basis of tunneling time definitions more general than the simple Wigner's phase-time \cite{Wig55} (Olkhovsky {\em et al.}, for instance, discuss a simple way of understanding the problem \cite{Olk04}).
In a {\em causal} manner, it might consist in explaining the superluminal phenomena during tunneling as simply due to a {\em reshaping} of the pulse, with attenuation, as already attempted (at the classical limit) \cite{Gav84}, i. e. the later parts of an incoming pulse are preferentially attenuated, in such a way that the outcoming peak appears shifted towards earlier times even if it is nothing but a portion of the incident pulse forward tail \cite{Ste93,Lan89}.
In particular, we do not intend to extend on the delicate question whether superluminal group-velocities can sometimes imply superluminal signalling, a controversial subject which has been extensively explored in the literature about the tunneling effect (\cite{Olk04} and references therein).

Turning back to the scattering time analysis, we can observe an analogy between our results and the results interpreted from the Hartman Effect (HE) analysis \cite{Har62}.
The HE is related to the fact that for opaque potential barriers the mean tunneling time does not depend on the barrier width, so that for large barriers the effective tunneling-velocity can become arbitrarily large, where it was found that the tunneling phase-time was independent of the barrier width.
It seems that the penetration time, needed to cross a portion of a barrier, in the case of a very long barrier starts to increase again after the plateau corresponding to infinite speed — proportionally to the distance \footnote{The validity of the HE was tested for all the other theoretical expressions proposed for the mean tunneling times \cite{Olk04}.}.
Our phase time dependence on the barrier width is similar to that which leads to Hartman interpretation as we can infer from Eqs.~(\ref{518}-\ref{519}).
Only when $\alpha$ tends to $0$ we have an explicit linear time-dependence on $L$ given by
\small\begin{equation}
t^{\varphi}_T = \frac{2\, m \, L}{w} \left(1 + \frac{1}{n}\right)
\label{14B}
\end{equation}\normalsize
which agree with calculations based on the simple phase-time analysis where $t_T = \frac{2\, m \, L}{w} \left(1 + \frac{1}{2n}\right)$.
However, it is important to emphasize that the wave packets for which we compute the phase times illustrated in the Fig.(\ref{fig3A}) are not {\em effectively} constructed with the same momentum distributions.
The phase $\Theta(k, L)$ appears when we treat separately each momentum amplitude $g(k - k_{\0})\,|T(k, L)|$ and $g(k - k_{\0})|R(k, L)|$ and the other one $\varphi(k, L)$ appears when we sum the amplitudes on the same side of the barrier.
The summed amplitude corresponds to a recomposed symmetrical momentum distribution for which the SPM will give the time-position of the peak of the corresponding wave packet.
In this sense, for overcoming the abovementioned incongruities, we have reconsidered the use of the multiple peak decomposition technique previously presented in the study of the above barrier diffusion problem \cite{Ber04}

\section{CONCLUSION}

In this manuscript we have discussed the wave packet multiple peak decomposition in order to identify some incongruities due to erroneous application of the method of stationary phase for deriving phase times related to wave packets scattered by a unidimensional potential barrier.
The main point on which we have elaborated shows us that the results of the SPM depend critically upon the manipulation of the amplitude prior to the application of the method.
The method is inherently ambiguous unless we know, by some other means, at least the number of separate peaks involved.
By analyzing the above barrier diffusion problem, we have demonstrated that the barrier results can be obtained by treating the barrier as a two-step process.
This procedure involves multiple reflections at each step and predict the existence of multiple (infinite) outgoing peaks.

In the same analysis which generically involves analytically-continuous {\em gaussian} pulses, holomorphic functions which do not have a well-defined front, we have quantified some aspects inherent to the tunneling phenomenon.
We have observed that the amplitude of the transmitted wave is essentially composed by the plane wave components of the front tail of the {\em incoming} wave packet which reaches the first barrier interface before the peak arrival.
As we have noticed, this {\em filter effect} by itself leads to ambiguities in the interpretation of the wave packet speed of propagation.
To overcome these incompatibilities which leads to the misconception of the tunneling phase time, we have considered
the multiple peak decomposition technique in a suitable way for comprehending the conservation of probabilities for a very particular tunneling configuration.
An example for which, we believe, we have provided a simple but convincing resolution, since the asymmetry presented in the previous case was eliminated, and the phase times could be accurately computed.

We let for a subsequent analysis the suggestive possibility of investigating the validity of our approach when confronting with the intriguing case of multiple opaque barriers \cite{Esp03}, in particular, in the case of non-resonant tunneling.
Still concerning with the future theoretical perspectives, the symmetrical colliding configuration also offers the possibility of exploring some applications involving soliton structures.
In summary, the above discussion reinforces the assertion that the investigation of wave propagation across a tunnel barrier has always been an intriguing subject which is wide open both from a theoretical and an experimental point of view.

\newpage

\end{document}